\documentclass[conference]{IEEEtran}
\usepackage{graphicx,times, amsmath, amsfonts}

\usepackage{amssymb,epstopdf}
\usepackage[noend]{algorithmic}
\usepackage{algorithm}
\ifCLASSINFOpdf
\else
\fi
\begin{document}

%===========================================================================
\title{Resource Allocation for Co-Primary Spectrum Sharing in MIMO Networks}
%============================================================================

%
\author{
\IEEEauthorblockN{Tachporn Sanguanpuak\IEEEauthorrefmark{1}, Sudarshan Guruacharya\IEEEauthorrefmark{2},
Nandana Rajatheva\IEEEauthorrefmark{1}, Matti Latva-aho\IEEEauthorrefmark{1}}
\IEEEauthorblockA{\IEEEauthorrefmark{1}Centre for Wireless Communications (CWC), University of Oulu, Finland; \\ \IEEEauthorrefmark{2}Dept. Elec. \& Comp. Eng., University of Manitoba, Canada}
\IEEEauthorblockA{Email: \{tachporn.sanguanpuak, nandana.rajatheva, matti.latva-aho\}@oulu.fi; sudarshan.guruacharya@umanitoba.ca}
}\maketitle

%%============================================================================
\begin{abstract}
%%============================================================================
We study co-primary spectrum sharing concept in two small cell multiuser network. Downlink transmission is explored with Rayleigh fading in interfering broadcast channel. Both base stations and all the users are equipped with multiple antennas. Resource allocation with joint precoder and decoder design is proposed for weighted sum rate (WSR) maximization problem. The problem becomes mixed-integer and non-convex. We factor the main objective problem into two subproblems. First subproblem is multiuser with subcarrier allocation where we assume that each subcarrier can be allocated to multiple users. Gale-Shapley algorithm based on stable marriage problem and transportation method are implemented for subcarrier allocation part. For the second subproblem, a joint precoder and decoder design is proposed to obtain the optimal solution for WSR maximization. Monte Carlo simulation is employed to obtain the results.

%%============================================================================
\end{abstract}
%%============================================================================

%%============================================================================
\section{Introduction}\label{section:intro}
%%============================================================================

Future wireless networks will have to satisfy the requirements and quality of service of a large amount of applications such as various information flows like video streaming, data apart from voice, and smart-phones etc. The number of devices in wireless network is increasing exponentially, and it could reach hundreds of billions within next decade. To satisfy a much larger low-rate devices and also high-rate mobile users, network operators will need to increase network capacity. The networks operators will need to allocate spectrum more efficiently to obtain higher network capacity.

Multi-operator spectrum sharing in which the multiple operators agree on jointly using parts of their licensed spectrum becomes a main contributor in this direction. This is mainly due to the fact that using of dedicated spectrum by operators has been shown to be inefficient, spectrum is found to be idle at various times. Multi-operator spectrum sharing concept is therefore expected to be an important aspect to improve spectral efficiency. In \cite{Marko2014}, licensed shared access (LSA) concept in $2.3$ GHz spectrum band is demonstrated for spectrum sharing between mobile network operator and incumbent users. The smart antennas technologies can be used to enhance LSA systems \cite{Seppo2014}. Different types of incumbent users and factors for allowing spectrum sharing with LSA from their perspective are considered in \cite{Miia2014}.

In \cite{Yong2014}, the authors consider co-primary spectrum sharing for dense small cells in which each operator divides spectrum pool into dedicated and shared spectrum band. System level simulation is used to evaluate the throughput. The spectrum sharing aspect in cognitive radio system is studied in \cite{Yuzhe2014}. The unlicensed secondary users try to maximize their overall capacity by cooperating with primary users. Sharing spectrum between the co-located radio networks (RANs) supported by different operators is studied in \cite{Sofonias2014}. A heuristic search algorithm is proposed to maximize the inter-RAN sum rate based on user-grouping, spectrum partitioning, and user scheduling.

There are a number of technology features and capabilities, which should be considered to be greatly beneficial in optimizing the utilization of available spectrum. For example, including multiple-input-multiple-output (MIMO), beamforming, and network densification.
By employing MIMO capability, the diversity gain and the coding gain can be improved through beamforming technique or space time coding compared to single-input-single-output (SISO) system \cite{Tarokh1998}. The joint precoder and decoder (known as a joint transceiver) shoud be designed properly to achieve higher throughput. Most of the MIMO broadcast channel (MIMO BC) works consider the linear joint transceiver techniques according to easily implementation \cite{WeiYu2007}, \cite{Wiesel2006}.

In \cite{Qingjiang2011}, multiple-BS transmit signal simultaneously to their users in their own cells and cause the interferences to users. A joint precoder and decoder design to maximize weighted sum rate (WSR) based on iterative minimization of weighted mean square error (WMSE) is proposed.

In 5th generation mobile communication (5G) system, multiple-input multiple-output (MIMO) and multi-operator spectrum sharing in heterogeneous network will play crucial roles. In the earlier works mentioned above, the concept of co-primary spectrum sharing in MIMO with multiple users networks has not been studied. In our work, we propose a co-primary spectrum sharing in multiuser two small cells network.
The small cell base stations and their own users employ multiple antennas. Joint precoder-decoder design based on MIMO system with subcarriers allocation for downlink transmission is proposed. We assume that both base stations allocate users in dedicated subbands and shared subbands. Each base station allocates its users to utilize the shared band when the number of subcarriers in dedicated spectrum band is not enough to serve all users.

The WSR maximization problem of both cells for multiple users with multiple subcarriers allocation is studied. The problem becomes non-convex and therefore we separate the main problem into two subproblems. In the first subproblem, we allocate subcarriers to users by using two methods. The first method is Gale-Shapley algorithm which is based on stable marriage problem \cite{Gale1962}. The second method is implemented based on transportation method \cite{NWflowbook}, linear programming is used to solve this problem. The number of users which is served by each subcarrier is less than or equal to the number of transmitting antennas for both methods. In the second subproblem, after a fixed subcarrier assignment, a joint precoder-decoder design is employed. Bisection method is used to find the optimal precoder.

%%============================================================================
\section{System Model} \label{section:BothBSsshareBW}
%%============================================================================
Both base stations allocate users in their dedicated bands first and then in the shared bands. In each cell, the base station allocates the same amount of bandwidth to each set of users.

\begin{figure}[h]
\centering
\includegraphics[height=8 in, width=3.4in, keepaspectratio = true]{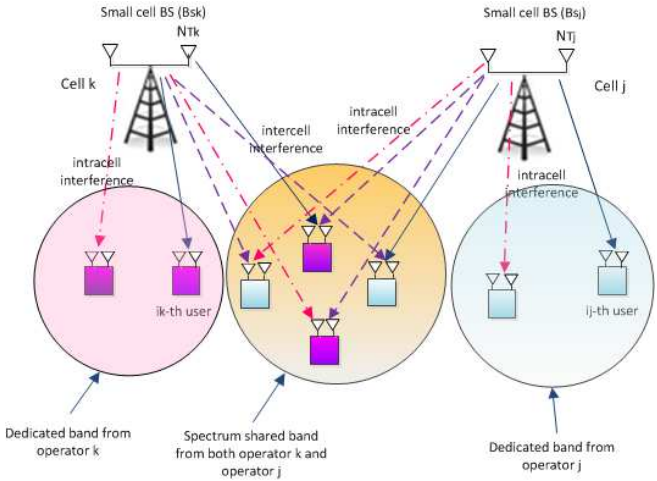}
\caption{Spectrum Sharing Between Two Small Cells}
\label{systemmodel}
\end{figure}

The base stations employ multiple number of antennas such that $N_{T_{k}}$ and $N_{T_{j}}$ for base station $k$ and $j$, where $j \neq k$, respectively. At both base stations, the number of total users is more than the number of transmit antennas denoted as, $I_{k} \geq N_{T_{k}}$ and $I_{j} \geq N_{T_{j}}$. $\mathcal{I}_{k}$, and $\mathcal{I}_{j}$ are the set of users in the cell $k$ and $j$ where $I_{k}$ is the cardinality of $\mathcal{I}_{k}$, denoted as $I_{k} = |\mathcal{I}_{k}|$. Each user in both cells utilize multiple number of antennas. Any user $i_{k}\in \mathcal{I}_{k}$ in the cell $k$ uses $N_{R_{k}}$ antennas and any user $i_{j}\in \mathcal{I}_{j}$ in the cell $j$ uses $N_{R_{j}}$ antennas. The system is demonstrated in Fig. \ref{systemmodel}.

The base station $k$ and $j$ allocate the set of subcarriers $\mathcal{N}$, and $\mathcal{M}$, respectively to their own users in the dedicated spectrum band. We denote the cardinality set of $\mathcal{N}$, and $\mathcal{M}$ as $N = |\mathcal{N}|$ and $M = |\mathcal{M}|$, respectively. In the cell $k$, and $j$ simultaneous transmission to multiple users on the same subcarrier $n$, and $m$, respectively is allowed.

Thus, in the dedicated band of the cell $k$, the received signal by user $i_{k}$ on subcarrier $n$ is given by,
\begin{align}
\mathbf{y}_{i_{k},n} &= \mathbf{H}_{i_{k},n}\mathbf{T}_{i_{k},n}\mathbf{x}_{i_{k},n} + \sum_{j_{k}=1,j_{k}\neq i_{k}}^{I_{k}}\mathbf{H}_{i_{k},n}\mathbf{T}_{j_{k},n}\mathbf{x}_{j_{k},n} \nonumber\\
&+ \mathbf{z}_{i_{k},n},
\end{align}
where $\sum_{j_{k}=1, j_{k}\neq i_{k}}^{I_{k}}\mathbf{H}_{i_{k},n}\mathbf{T}_{j_{k},n}\mathbf{x}_{j_{k},n}$ is the intracell interference (ICI).
The $\mathbf{H}_{i_{k},n}$ is $N_{R_{k}} \times N_{T_{k}}$ channel gain matrix between the base station $k$ and any user $i_{k}$ on subcarrier $n$.
$\mathbf{T}_{i_{k},n}$ is the $N_{T_{k}} \times a_{i_{k}}$ beamforming matrix allocated to user $i_{k}$ on subcarrier $n$ and $a_{i_{k}}$ denotes independent data streams from the base station $k$ to user $i_{k}$. $\mathbf{x}_{i_{k},n}$ is the column matrix of length $a_{i_{k}} \times 1$ representing transmitted symbols to different $I_{k}$ users. The $\mathbf{z}_{i_{k},n}$ is additive white Gaussian noise with zero mean and covariance matrix $\sigma_{i_{k},n}^{2}\mathbf{I}_{N_{R_{k}}}$.

It is required to select $t_{k} \leq N_{T_{k}}$ users out of the set of $\mathcal{I}_{k}$ users in each subcarrier. Number of simultaneously served users on each subcarrier is limited by the number of transmit antennas. There are $C_{k} = \sum_{t_{k} =1}^{N_{T_{k}}}\binom{I_{k}}{t_{k}}$ combinations of users who can utilize same subcarrier, each of them is denoted as $\mathcal{L}_{t_{k}}$, where $\mathcal{L}_{t_{k}} = \{1,...,I_{k}\}$, $0 < |\mathcal{L}_{t_{k}}| \leq N_{T_{k}}$. Assuming that the group of users $\mathcal{L}_{t_{k}}$ is assigned to subcarrier $n$, then received signal by any user ($i_{k} \in \mathcal{L}_{t_{k}}$) is as, $\mathbf{y}_{i_{k},n} = \mathbf{H}_{i_{k},n}\mathbf{T}_{i_{k},n}\mathbf{x}_{i_{k},n} + \sum_{j_{k}\in \mathcal{L}_{t_{k}}, j_{k}\neq i_{k}}\mathbf{H}_{i_{k},n}\mathbf{T}_{j_{k},n}\mathbf{x}_{j_{k},n}+ \mathbf{z}_{i_{k},n},$
where $\sum_{j_{k}\in \mathcal{L}_{t_{k}}, j_{k}\neq i_{k}}\mathbf{H}_{i_{k},n}\mathbf{T}_{j_{k},n}\mathbf{x}_{j_{k},n}$ denotes the interference caused by users on the same subcarrier $n$.

In the shared band part, both intracell and intercell interference affect all the users. The intracell interference is caused by the same subcarrier allocated to users and the intercell interference is caused by the broadcast signal from the other base station in the same frequency for shared band.

At the receiver, a decoding matrix $\mathbf{U}_{i_{k},n}$ with the size $N_{R_{k}} \times a_{i_{k}}$ is used to recover the original signal. We assume that the receiver use MMSE decoder thus, the $\mathbf{U}_{i_{k},n}$ denotes MMSE decoding matrix from the base station $k$ to user $i_{k}$ on subcarrier $n$. The output of the decoder can be written as $\mathbf{r}_{i_{k},n} = \mathbf{U}_{i_{k},n}^{H}\mathbf{y}_{i_{k},n}$. In the cell $j$, we consider the same scenario as in the cell $k$. Thus, we can write the output of the decoder in the cell $j$ similar to cell $k$ as  $\mathbf{r}_{i_{j},m} = \mathbf{U}_{i_{j},m}^{H}\mathbf{y}_{i_{j},m}$.

The maximization of total weighted sum rate of two cells in both dedicated band and shared band parts are studied in next section.

%%============================================================================
\section{Weighted Sum Rate Maximization for Co-Primary Spectrum Sharing} \label{section:BothBSsshareBW}
%%============================================================================
The optimization problem for maximization weighted sum rate of both cells can be written as,
\begin{eqnarray}
\text{max} & \sum_{i_{k}=1}^{I_{k}}\sum_{n=1}^{N}\mu_{i_{k},n}R_{k} + \sum_{i_{j}=1}^{I_{j}}\sum_{m=1}^{M}\mu_{i_{j},m}R_{j}\nonumber\\
\label{powconstrainorigi}
\text{subject to} & \sum_{i_{k} \in I_{k}}\sum_{n \in \mathcal{N}}||\mathbf{T}_{i_{k},n}||^{2}\leq P_{k}^{max}, \nonumber\\
&\sum_{i_{j} \in \mathcal{I}_{j}}\sum_{m \in \mathcal{M}}||\mathbf{T}_{i_{j},m}||^{2}\leq P_{j}^{max} \nonumber\\
& \sum_{i_{k}\in\mathcal{I}_{k}}\rho_{i_{k},n,t_{k}} \leq N_{T_{k}} \quad \forall n,t_{k} \nonumber\\
& \sum_{i_{j}\in\mathcal{I}_{j}}\rho_{i_{j},m,t_{j}} \leq N_{T_{j}} \quad \forall m,t_{j} \nonumber\\
&\rho_{i_{k},n,t_{k}}, \rho_{i_{j},m,t_{j}} \in \{0,1\} \quad \forall n,m,t_{k},t_{j},
\label{mainoptprob1}
\end{eqnarray}
where $R_{k} = \sum_{t_{k}=1}^{C_{k}}\rho_{i_{k},n,t_{k}}R_{i_{k},n}$, where $R_{i_{k},n}$ is the rate of user $i_{k}$ on subcarrier $n$. And $R_{j} = \sum_{t_{j}=1}^{C_{j}}\rho_{i_{j},m,t_{j}}R_{i_{j},m}$, where $R_{i_{j},m}$ is the rate of user $i_{j}$ on subcarrier $m$. $\rho_{i_{k},n,t_{k}}$ denotes whether subcarrier $n$ is assigned to user $i_{k} \in \mathcal{L}_{t_{k}}$ in the cell $k$, then $\rho_{i_{k},n,t_{k}} = 1$, otherwise $\rho_{i_{k},n,t_{k}} = 0$.  And $\rho_{i_{j},m,t_{j}}$ denotes whether subcarrier $m$ is assigned to user $i_{j} \in \mathcal{L}_{t_{j}}$ in the cell $j$, then $\rho_{i_{j},m,t_{j}} = 1$, otherwise $\rho_{i_{j},m,t_{j}} = 0$. We assume that each particular subcarrier is allocated to more than one user. In addition the number of users which are served by each subcarrier is less than or equal to number of transmitting antennas. $\mu_{i_{k},n}$ and $\mu_{i_{j},m}$ are weights which are used to represent the priority of any user $i_{k}$ and $i_{j}$ in the cell $k$ and $j$, respectively.

This problem is mixed-integer and nonconvex, we propose a heuristic method to solve (\ref{mainoptprob1}). To reduce the computational complexity, first we find subcarrier assignment and in the second stage, we find the precoder for users in the particular subcarrier by using fixed subcarrier assignment. The algorithm for subcarrier allocation in the cell $k$ is described as follows according to Gale-Shapley algorithm.\cite{Gale1962} Essentially, Gale-Shapley algorithm solves a matching problem (also known as the stable marriage problem) for a bipartite graph, where the two disjoint set of vertices are customarily referred to as the set of men and set of women. A man and a woman is said to have a stable marriage if neither of them are better off being with a different partner. The basic idea behind the algorithm is to allow men to propose to women in the order of their preference, while women provisionally accepts the proposal until a better proposal arrives. Unlike the assignment problems solved by the Hungarian method for weighted bipartite graph, where the objective is to obtain a maximum weighted matching, the focus of Gale-Shapley method is to obtain a stable matching.

In our case, we consider a slight modification of the stable marriage problem, in that a single woman can engage with more than one man. We assume that the role of men is assumed by the users, while the role of women is assumed by the subcarriers. However, our modification of the basic Gale-Shapley algorithm allows a subcarrier to accept more than one user. If the number of subcarriers is not enough to serve all of it's users, the base station will assign the remaining users to utilize the shared band. Note that in the cell $j$, we repeat the same process as in the cell $k$. The algorithm for subcarrier allocation with joint precoder and decoder designs is described in Table I. The worst case complexity of this algorithm is $\mathcal{O}((|\mathcal{N}|+|\mathcal{M}|)\sum_k I_k)$.

\begin{table}[h]
\begin{tabular}{|r l|}
\hline
1.& Set $R_{i_{k}}=0$ for $i_{k} \in \mathcal{I}_{k}$ \\
2.& For each user $i_{k} \in \mathcal{I}_{k}$, in cell $k$, make a preference list of dedicated\\
  & subchannels $\mathcal{B}_{i_{k}} = [\alpha_1,...,\alpha_N]$ such that\\
  &$||\mathbf{H}_{i_{k},\alpha_l}||^{2} \geq ||\mathbf{H}_{i_{k},\alpha_m}||^{2}$ if $l<m$\\
3.& For each dedicated subchannel $n\in\mathcal{N}$, make a preference list \\
  &of users $\mathcal{Q}_n = [\beta_1,...,\beta_N]$ such that\\
  &$||\mathbf{H}_{\beta_l,n}||^{2} \geq ||\mathbf{H}_{\beta_m,n}||^{2}$ if $l<m$\\
4.& For each subchannel $n\in\mathcal{N}$, initialize the user acceptance list \\
  &$\mathcal{A}_n = \emptyset$ \\
5.& Repeat \\ %For t = 1 to N,  \\
6.& \quad For each user $i_{k} \in \mathcal{I}_{k}$, \\
7.& \qquad If $\nexists n$ such that $i_{k} \in \mathcal{A}_n$, (i.e. if there is no acceptance list\\
  & \qquad to which user $i_{k}$ belongs to)\\
8.& \qquad \quad Find the subchannel $\alpha_t \in \mathcal{B}_i$ with the highest preference, \\
9.& \qquad \quad If $|\mathcal{A}_{\alpha_t}| < N_T$, \\
10.& \qquad \qquad Put user $i_{k}$ in the acceptance list $\mathcal{A}_{\alpha_t}$ \\
11.& \qquad \quad Else if $\exists \gamma \in \mathcal{A}_{\alpha_t}$ such that $\beta^{-1}(i)<\beta^{-1}(\gamma)$ in the\\
   & \qquad \quad preference list $\mathcal{Q}_{\alpha_t}$ \\
12.& \qquad \qquad Replace the user $\gamma$ by user $i$ in $\mathcal{A}_{\alpha_t}$ \\
13.& \qquad \quad End If \\
14.& \qquad Remove $\alpha_t$ from the preference list $\mathcal{B}_i$ \\
15.& \qquad End If \\
16.& \quad End For \\
17.& Until \{$\exists n$ such that $i_{k}\in\mathcal{A}_n$ for every $i_{k} \in \mathcal{I}_{k} \}$ OR \\
   &\{$\mathcal{B}_{i_{k}} = \emptyset$ for any $i_{k} \in \mathcal{I}_{k}$ \}\\
18.& Find the subset of users $\mathcal{I}_{k}' \subset \mathcal{I}_{k}$ which have not been assigned \\
   &to any acceptance list \\
19.& Do steps 2 to 17 for users in $\mathcal{I}_{k}'$ using shared subchannels. \\
20.& For each user $i_{k} \in \mathcal{I}_{k}$,\\
21.& \quad Allocate power in that particular subchannel $n$ based on \\
   & \quad optimal joint transceiver design via WMMSE \\
   & \quad iterative algorithm\\
22.& End For\\
23.& For each user $i_{k} \in\mathcal{I}_{k}$,\\
24.& \quad Compute the rate achieved $R_{k}$ \\
25.& End For \\
\hline
\end{tabular}
\caption{Subcarriers allocation by using Gale-Shapley algorithm with joint precoder-decoder design}
\label{subcarrierallocate:alg1}
\end{table}

%%============================================================================
\section{Weighted Sum Rate Maximization via Weighted sum MSE Minimization for Co-Primary Spectrum Sharing} \label{section:WMSEmin}
%%============================================================================
In this section, we study a joint precoder-decoder design in MIMO system for the second stage subproblem.
Due to the equivalence between WSR maximization problem and minimization of WMSE problem \cite{Qingjiang2011}, we can obtain the optimal solution of WSR maximization via WMSE minimization. In the dedicated band of the cell $k$, the MSE covariance matrix between the actual transmitted data vector and the received signal at any user $i_{k}$ on the subcarrier $n$ can be written as $\text{MSE}(\mathbf{T}_{i_{k},n}, \mathbf{U}_{i_{k},n}) = \mathbb{E}_{\mathbf{x}_{i_{k},n},\mathbf{z}_{i_{k},n}}[||\mathbf{x}_{i_{k},n} - \mathbf{U}^{H}_{i_{k},n}\mathbf{y}_{i_{k},n}||^{2}]$ under the assumption that $\mathbf{x}_{i_{k},n}$ and $\mathbf{z}_{i_{k},n}$ are independent. It leads to,
\begin{align}
&\text{MSE}(\mathbf{T}_{i_{k},n}, \mathbf{U}_{i_{k},n})\nonumber\\
&= \mathbb{E}_{\mathbf{x}_{i_{k},n},\mathbf{z}_{i_{k},n}}[(\mathbf{x}_{i_{k},n} - \mathbf{U}^{H}_{i_{k},n}\mathbf{y}_{i_{k},n})(\mathbf{x}_{i_{k},n} - \mathbf{U}^{H}_{i_{k},n}\mathbf{y}_{i_{k},n})^{H}]\nonumber\\
& = (\mathbf{I}_{N_{R_{k}}}- \mathbf{U}_{i_{k},n}^{H}\mathbf{H}_{i_{k},n}\mathbf{T}_{i_{k},n})(\mathbf{I}_{N_{R_{k}}} - \mathbf{U}_{i_{k},n}^{H}\mathbf{H}_{i_{k},n}\mathbf{T}_{i_{k},n})^{H} \nonumber\\
&+  \sum_{j_{k}\in \mathcal{L}_{t_{k}}, j_{k}\neq i_{k}}\mathbf{U}_{i_{k},n}\mathbf{H}_{i_{k},n}\mathbf{T}_{j_{k},n}\mathbf{T}_{j_{k},n}^{H}\mathbf{H}_{i_{k},n}^{H}\mathbf{U}_{i_{k},n}^{H} \nonumber\\
& + \sigma_{i_{k},n}^{2}\mathbf{U}_{i_{k},n}^{H}\mathbf{U}_{i_{k},n}
\label{MSEmatrix}
\end{align}

Additionally, in the dedicated band of the cell $j$, the MSE covariance matrix at any user $i_{j}$ on the subcarrier $m$ denoted as $\text{MSE}(\mathbf{T}_{i_{j},m}, \mathbf{U}_{i_{j},m})$ can be written equivalent to $\text{MSE}(\mathbf{T}_{i_{k},n}, \mathbf{U}_{i_{k},n})$.
In the shared band of the cell $k$, the MSE covariance matrix of any remaining user $l_{k}$ which has not been assigned to the dedicated band can be given as,
\begin{align}
&\text{MSE}(\mathbf{T}_{l_{k},p}, \mathbf{U}_{l_{k},p})\nonumber\\
& = (\mathbf{I}_{N_{R_{k}}}- \mathbf{U}_{l_{k},p}^{H}\mathbf{H}_{l_{k},p}\mathbf{T}_{l_{k},p})(\mathbf{I}_{N_{R_{k}}} - \mathbf{U}_{l_{k},p}^{H}\mathbf{H}_{l_{k},p}\mathbf{T}_{l_{k},p})^{H} \nonumber\\
&+  \sum_{k=1}^{K}\sum_{b_{k}\in \mathcal{L}_{t_{k}\text{share}}, b_{k}\neq l_{k}}\mathbf{U}_{l_{k},p}\mathbf{H}_{l_{k},p}\mathbf{T}_{b_{k},p}\mathbf{T}_{b_{k},p}^{H}\mathbf{H}_{b_{k},p}^{H}\mathbf{U}_{l_{k},b}^{H} \nonumber\\
& + \sigma_{l_{k},p}^{2}\mathbf{U}_{l_{k},p}^{H}\mathbf{U}_{l_{k},p}
\label{MSEmatrixshareband}
\end{align}
where $p$ is the subcarrier in shared band which any remaining user $l_{k}$ is assigned. The MSE covariance matrix of any remaining user $l_{j}$ of the cell $j$ which is assigned to shared band can be written equivalent to $\text{MSE}(\mathbf{T}_{l_{k},p}, \mathbf{U}_{l_{k},p})$.
%In addition, the MSE covariance matrix at any user $i_{j}$ on the subcarrier $m$ denoted as $\text{MSE}(\mathbf{T}_{i_{j},m}, \mathbf{U}_{i_{j},m})$ %can be written equivalent to $\text{MSE}(\mathbf{T}_{i_{k},n}, \mathbf{U}_{i_{k},n})$.
With an extension of \cite{Qingjiang2011}[Theorem 1], for $\mathbf{W}_{i_{k},n} \geq 0$ be a weight matrix at any user $i_{k}$ on the subcarrier $n$. The following problem has the same optimal solution as weighted sum rate maximization problem and can be rewritten as,
\begin{eqnarray}
\text{min} & \sum_{i_{k}=1}^{I_{k}}\sum_{n=1}^{N}\mu_{i_{k},n}F_{i_{k},n}\nonumber\\
           &  +\sum_{i_{j}=1}^{I_{j}}\sum_{m=1}^{M}\mu_{i_{j},m}F_{i_{j},m}\nonumber\\
\label{powconstrain}
\text{subject to} & \sum_{i_{k} \in I_{k}}\sum_{n \in \mathcal{N}}||\mathbf{T}_{i_{k},n}||^{2}\leq P_{k}^{max}, \nonumber\\
&\sum_{i_{j} \in \mathcal{I}_{j}}\sum_{m \in \mathcal{M}}||\mathbf{T}_{i_{j},m}||^{2}\leq P_{j}^{max}
\label{mainoptprob2}
\end{eqnarray}
where
\begin{align}
F_{i_{k},n} &= \text{Tr}(\mathbf{W}_{i_{k},n}\times\text{MSE}(\mathbf{T}_{i_{k},n}))-\text{log}_{2}\text{det}(\mathbf{W}_{i_{k},n})\\
F_{i_{j},m} &= \text{Tr}(\mathbf{W}_{i_{j},m}\times\text{MSE}(\mathbf{T}_{i_{j},m}))-\text{log}_{2}\text{det}(\mathbf{W}_{i_{j},m})
\end{align}
For fixed receivers which are given by the MMSE solution, it leads to,
\begin{align}
\mathbf{U}_{i_{k},n}^{\text{MMSE}} = \mathbf{J}_{i_{k},n}^{-1}\mathbf{H}_{i_{k},n}\mathbf{T}_{i_{k},n},
\label{UMMSE}
\end{align}
$\mathbf{J}_{i_{k},n}= \sum_{j_{k}=1,j_{k}\neq i_{k}}^{\mathcal{I}_{k}}\sum_{n=1}^{N}\mathbf{H}_{i_{k},n}\mathbf{T}_{j_{k},n}\mathbf{T}_{j_{k},n}^{H}\mathbf{H}_{i_{k},n}^{H} + \sigma_{i_{k},n}^{2}\mathbf{I}$.
With the MMSE receiver $\mathbf{U}_{i_{k},n}^{\text{MMSE}}$, the following MSE matrix is given by
\begin{align}
\text{MSE}(\mathbf{T}_{i_{k},n}, \mathbf{U}_{i_{k},n}^{\text{MMSE}}) = \mathbf{I}-\mathbf{T}_{i_{k},n}\mathbf{H}_{i_{k},n}\mathbf{J}_{i_{k},n}^{-1}\mathbf{H}_{i_{k},n}\mathbf{T}_{i_{k},n}
\end{align}

The coordinate descent algorithm based on \cite{Qingjiang2011} is used to solve (\ref{mainoptprob2}). The coordinated descent algorithm is performed by three sets of variables which are precoder, decoder, and weight matrices. Each set of these three variables is solved seperately in a sequential manner, while assuming the other two sets are fixed. Thus, we update the transmit beamformer $\mathbf{T}_{i_{k},n}$ and weight $\mathbf{W}_{i_{k},n}$ by fixing the decoder matrix.
%%The updating weight $\mathbf{W}_{i_{k},n}$ is obtained by using gradient of objective function as $\mathbf{W}_{i_{k},n} = ( \text{MSE}(\mathbf{T}_{i_{k},n}, \mathbf{U}_{i_{k},n}^{\text{MMSE}}))^{-1}$.

The Karush-Kuhn-Tucker (KKT) condition can be used to find the optimal solution of the problem (\ref{mainoptprob2}). With the first order optimality condition of Lagragian function respect to $\mathbf{T}_{i_{k},n}$, this yields
\begin{align}
\mathbf{T}_{i_{k},n}^{\text{opt}} &= \mu_{i_{k},n}\bigg(\sum_{j_{k}=1,j_{k}\neq i_{k}}^{\mathcal{I}_{k}}\sum_{n=1}^{N}\mathbf{H}_{j_{k},n}^{H}\mathbf{U}_{j_{k},n}^{\text{MMSE}}\mathbf{W}_{j_{k},n}\nonumber\\
&\times (\mathbf{U}_{j_{k},n}^{\text{MMSE}})^{H}\mathbf{H}_{j_{k},n}+ \lambda_{i_{k},n}^{*}\mathbf{I}\bigg)^{-1}\mathbf{H}_{i_{k},n}^{H}\mathbf{U}_{i_{k},n}^{\text{MMSE}}\mathbf{W}_{i_{k},n}
\label{Tiknopt}
\end{align}
where the optimum $\lambda_{i_{k},n}^{*}$ must be positive and it can be obtained by using the bisection method. To obtain the joint precoder and decoder design for WMSE minimization after subcarrier assignment, an iterative algorithm is proposed in Table II.
Note that the data rate of each user $i_{k}$ on the subcarrier $n$ in the dedicated band of cell $k$ can be obtain as $R_{i_{k},n}=\text{log}_{2}\text{det}((\text{MSE}(\mathbf{T}_{i_{k},n}, \mathbf{U}_{i_{k},n}^{\text{MMSE}}))^{-1})$. All the users�s data rate are calculated similar as $R_{i_{k},n}$.

\begin{table}[ht]
\begin{tabular}{|r l|}
\hline
1.& Initialize the precoding matrix $\mathbf{T}_{i_{k},n}$ such that $\text{Tr}(\mathbf{T}_{i_{k},n}\mathbf{T}_{i_{k},n}^{H}) = \frac{P_{t}}{I_{K}}$ \\
2.& Repeat \\
3.& $\mathbf{U}_{i_{k},n} \leftarrow \mathbf{U}_{i_{k},n}^{\text{MMSE}} =   \mathbf{J}_{i_{k},n}^{-1}\mathbf{H}_{i_{k},n}\mathbf{T}_{i_{k},n}$ in (\ref{UMMSE})\\
4.& $\mathbf{W}_{i_{k},n} \leftarrow (\mathbf{I} - \mathbf{U}_{i_{k},n}^{H}\mathbf{H}_{i_{k},n}\mathbf{T}_{i_{k},n})^{-1}$\\
5.& Find the optimum Lagrange multiplier value $\lambda_{i_{k},n}^{*}$ by using \\
  & bisection method \\
6.& Substitute the optimum $\lambda_{i_{k},n}^{*}$ in $\mathbf{T}_{i_{k},n}^{\text{opt}}$ in (\ref{Tiknopt}) \\
7.& Put $\mathbf{T}_{i_{k},n} \leftarrow \mathbf{T}_{i_{k},n}^{\text{opt}}$ in (\ref{Tiknopt})\\
8.& Untill $|(\text{log}_{2}\text{det}(\mathbf{W}_{i_{k},n}))^{b+1} - (\text{log}_{2}\text{det}(\mathbf{W}_{i_{k},n}))^{b}| \leq \epsilon$ where $b$  \\
  & denotes the iteration number and $\epsilon$ is a tolerence value\\
  & ($0 < \epsilon << 1$)\\
\hline
\end{tabular}
\caption{An iterative algorithm for a joint precoder and decoder design via WMMSE after subcarrier allocation based on Gale-Shapley algorithm in the cell $k$}
\label{Sequentialupdat:alg1}
\end{table}

In the cell $j$, an iterative algorithm to obtain the optimal precoder and decoder design can be proposed similar to as in the cell $k$.

%%============================================================================
\section{Numerical Results}\label{section:Results}
%%============================================================================
The numerical results illustrate performance of maximization of WSR via WMSE minimization problem for MIMO with co-primary spectrum sharing. Rayleigh fading channels are assumed in MIMO interfering broadcast channels. The uncorrelated fading channels are generated for each user equipment independently in each time slot.

Monte Carlo simulation is used by assuming $5000$ samples to obtain the results. After both base stations finish allocating subcarriers to their users in the dedicated band there may still be some remaining users. Both base stations will assign those users to employ shared band.
In all the simulation, we assume that the number of antennas at both base stations $N_{T_{k}}, N_{T_{j}} = 4$ and at each user, $N_{R_{k}}, N_{R_{j}} = 2$

To compare the performance of Gale-Shapely method, we model the subchannel allocation problem as a transportation problem, which is a special case of maximum weight flow problem, which is then solved by using linear programming. The transportation problem itself is a generalization of the assignment problem which can be solved using the Hungarian method. The Hungarian method is not applicable in our case since a subchannel can be assigned to more than one user. In a nutshell, we have a bipartite graph where the subchannels are source nodes with outflow constraint equal to the maximum number of users they can support, which is equal to the number of transmit antennas. Similarly, the destination nodes are the users with inflow constraint equal to unity, that is, a user can be assigned to a single subchannel. Lastly, the weight matrix is given by the channel gain matrix.
.

\begin{figure}[h]
\centering
\includegraphics[height=5 in, width=3.6in, keepaspectratio = true]{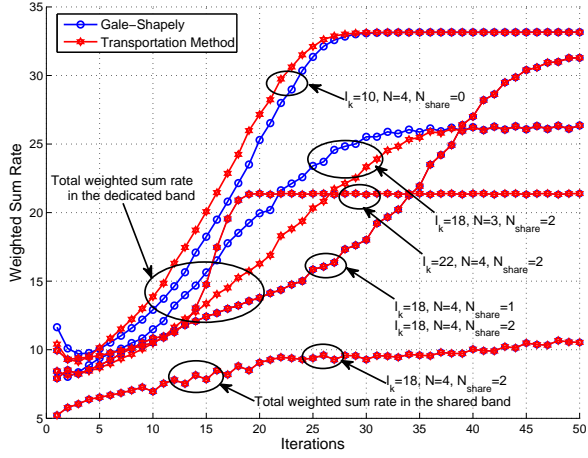}
\caption{Weighted sum rate versus iterations for MIMO co-primary spectrum sharing}
\label{WSRfig1}
\end{figure}

Fig.2. demonstrates the overall WSR of both small cells. We show the overall WSR in the dedicated band and shared band of both cells versus iterations. We assume that the number of users in both cells ($I_{k}$ and $I_{j}$) is equal ($I_{k}=I_{j}=10, 18$) for each case. The number of subcarriers for dedicated band in the cell $k$ and $j$ is $N=M=3,4$. The number of subcarriers for shared band $N_{\text{share}}=M_{\text{share}}=0,1,2$ in both cells. It can be observed that when $I_{k} =18, N=3, N_{\text{share}}=2$, the Gale-Shapley converge faster than the transportation method. When $I_{k}=10, N=4, N_{\text{share}}=0$, the transportation method converges little faster. In most cases, the Gale-Shapley and transportation method give similar WSR values versus iterations and converge the same time.

\begin{figure}[h]
\centering
\includegraphics[height=5.2 in, width=3.5in, keepaspectratio = true]{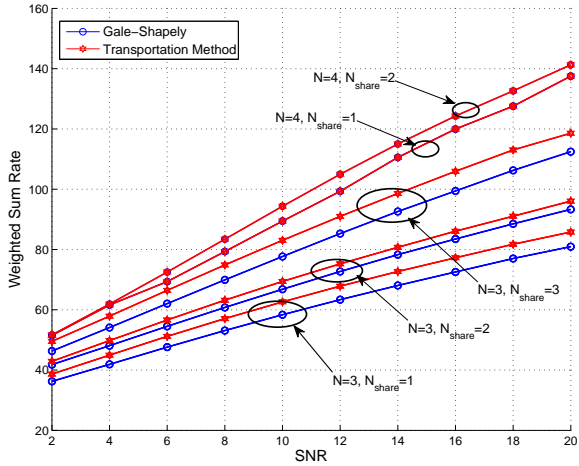}
\caption{Weighted sum rate versus SNR for MIMO co-primary spectrum sharing }
\label{WSRfig3}
\end{figure}

Fig.3. demonstrates the the overall WSR in both cells versus SNR. We assume that the number of users in both cells ($I_{k}$ and $I_{j}$) is equal ($I_{k}=I_{j}=18$) for each case. We assume that the number of subcarriers for dedicated band in the cell $k$ and $j$ $\mathcal{N}=\mathcal{M}=3,4$ and the number of subcarriers for shared band $\mathcal{N}_{\text{share}}=\mathcal{M}_{\text{share}}=1,2,3$ in both cells. When the number of subcarriers in shared band is increased, the weighted sum rate improved for both Gale Shapley and transportation methods. For $N=3, N_{\text{share}}=1,2,3$ the transportation method gives higher WSR throughput. But for $N=4, N_{\text{share}}=1,2$ both Gale-Shapley and transportation methods give the same values.

The WSR is enhanced with increasing number of subcarriers in the shared band according to the number of users is more than number of transmit antennas multiplied by number of dedicated subcarriers thus, both base stations assign the remaining users to utilize their shared band. Although, intercell interference is available at remaining users allocated in the shared band.
%In addition, when number of transmit antennas $N_{T_{k}}, N_{T_{j}}$ increases, the weighted sum rate is enhanced.

%
\begin{figure}[h]
\centering
\includegraphics[height=5 in, width=3.6in, keepaspectratio = true]{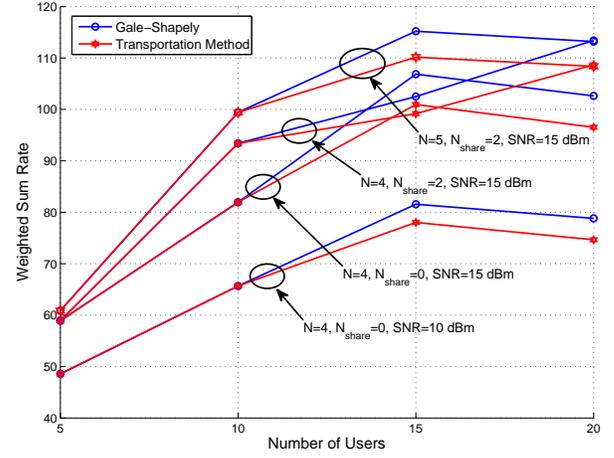}
\caption{Weighted sum rate versus number of users for MIMO co-primary spectrum sharing }
\label{WSRfig3}
\end{figure}

Fig.4. demonstrates the the overall WSR in both cells versus number of users with the fixed SNR values. We illustrate the total WSR of both cells by varying with number of users with $\text{SNR} =10, 15$ dBm. The number of subcarriers for dedicated band in the cell $k$ and $j$ $\mathcal{N}=\mathcal{M}=4,5$. The number of subcarriers for shared band $\mathcal{N}_{\text{share}}=\mathcal{M}_{\text{share}}=0,2$. We can observe that with increasing the values of SNR, the WSR is improved. When the number of users is from $5$ to $10$, the Gale-Shapley and transportation method give similar values. For higher number of users such that from $10$ to $20$ users, the Gale-Shapley provides higher throughput for all cases.

%%============================================================================
\section{Conclusion}\label{section:conclusion}
%%============================================================================
We propose co-primary spectrum sharing concept with the MIMO multiuser for two small cell network in downlink transmission. The overall weighted sum rate maximization of both small cells is studied with subcarrier allocation and a joint precoder and decoder design.
The weighted sum rate maximization problem becomes mixed-integer and non-convex problem. We separate the main optimization into two sub-problems. In the first subproblem, we assume that both base stations allocate the subcarriers to each user by employing two methods. The first method is Gale-Shapley based on stable marriage problem and the second one is based on transportation problem. The linear programming is used to solve the transportation problem. In both methods, each subcarrier can be allocated to multiple users. In the second sub-problem, a joint precoder-decoder design for MIMO multiuser is proposed. The maximization of weighted sum rate is solved via weighted mean square error minimization problem for a joint transceiver design. The optimal precoder is obtained by the bisection method.

Numerical results illustrate the over all weighted sum rate for both base stations with uncorrelated antennas. We show the convergence of the Gale-Shapley and the transportation problem. The overall weighted sum rate throughput is demonstrated for different number of subcarriers served in dedicated and shared bands for both cells. Then, with different number of users and fixed the value of SNR, the overall weighted sum rate is explored. In terms of the convergence, these two methods provide similar behavior. For the WSR versus number of users, the Gale-Shapley gives higher WSR throughput than the transportation problem for high number of users. When the number of subcarriers in the dedicated band is not enough to serve all users, both base stations will assign the remaining users to utilize the shared band. The overall average weighted sum rate is improved significantly in this instance. In terms of the complexity, the transportation method has leads more complexity because the linear programming is used to solved the problem. Thus, the Gale-Shapley is a good method for multiple subcarriers with multiple users assignment problem.

%%%=======================================================================
\bibliographystyle{IEEE}

\begin{thebibliography}{1}
%%%=======================================================================

\bibitem{Marko2014}
M.~Palola, {\em et al.}, ``Licensed Shared Access (LSA) trial demonstration using real LTE network,'' {\em Proc. Crowncom}, 2014. 

\bibitem{Seppo2014}
S.~Yrjola and E.~Heikkinen, ``Active Antenna System Enhancement for Supporting Licensed Shared Access (LSA) Concept,'' {\em Proc. Crowncom}, 2014. 

\bibitem{Miia2014}
M.~Mustonen, {\em et al.}, ``Spectrum Sharing and Energy-Efficient Power Optimization for Two-tier Femtocell Networks, '' {\em Proc. Crowncom}, 2014.

\bibitem{Yong2014}
Y.~Teng, Y.~Wang and K~.~Horneman, ``Co-Primary Spectrum Sharing for Denser Networks in Local Area, '' {\em Proc. Crowncom}, 2014

\bibitem{Yuzhe2014}
Y.~Xu, L.~Wang, C.~Fischione, and V.~Fodor, ``Distributed spectrum leasing via vertical cooperation in spectrum sharing networks, '' {\em Proc. Crowncom}, 2014. 


\bibitem{Sofonias2014}
S.~Hailu, A.~A.~Dowhuszko, O.~Tirkkonen, ``Adaptive Co-primary Shared Access Between Co-located Radio Access Networks, '' {\em Proc. Crowncom}, 2014. 


\bibitem{Tarokh1998}
V.~Tarokh, N.~Seshadri and A.~Calderbank, ``Space-time codes for high data rate wireless communication: performance criterion and code construction,'' {\em IEEE Trans on Information Theory}, vol.44, no. 2, pp.744-765, 1998. 


\bibitem{WeiYu2007}
W.~Yu and T.~Lan, ``Transmitter Optimization for the Multi-Antenna Downlink With Per-Antenna Power Constraints,'' {\em IEEE Trans on Signal Processing}, vol. 55, no. 6, pp. 2646-2660, 2007.


\bibitem{Wiesel2006}
A.~Wiesel, Y.-C.~Eldar, and S.~Shamai, ``Linear precoding via conic optimization for fixed MIMO receivers,'' {\em IEEE Trans on Signal Processing}, vol. 54, no. 1, pp. 161-176, 2006.


\bibitem{Qingjiang2011}
Q.~Shi, M.~Razaviyayn, Z.-Q.~Luo, and C.~He, ``An Iteratively Weighted MMSE Approach to Distributed Sum-Utility Maximization for a MIMO Interfering Broadcast Channel,'' {\em IEEE Trans. on Signal Processing}, vol. 59, no. 9, pp. 4331-4340, 2011. 


\bibitem{Gale1962}
D.~Gale, and L.S.~Shapley, ``College Admissions and Stability of Marriage,'' {\em The American Mathematical Monthly Jstor}, vol.69, no. 1, pp. 9-15, Jan. 1962.

\bibitem{NWflowbook}
K.-A.~Ravindra, T.-L.~Magnanti, and J.-B.~Orlin, ``Network Flows: Theory, Algorithms and Applications,'' {\em Prentice Hall}, 1993. 



%
%
%@ARTICLE{Ishtiaq2014,
%author={Ahmad, I. and Feng, Z. and Hammed, A. and Zhang, P. and Zhao, Y.},
%journal={in Proc. Crowncom},
%title={Spectrum Sharing and Energy-Efficient Power Optimization for Two-tier Femtocell Networks},
%year={2014},
%}
%
%
%@BOOK{Boydconvexbook2004,
%  title = {Convex Optimization},
%  publisher = {New York, NY: Cambridge University Press},
%  year = {2004},
%  author = {Boyd, Stephen and Vandenberghe, Lieven},
%}
%
%
%@BOOK{Bertsekas1999,
%  title = {Nonlinear Programming},
%  publisher = {Belmont, MA : Athena Scientific},
%  year = {1999},
%  author = {Bertsekas, D P.},
%}



%@ARTICLE{Eleftherios2011,
%author={Karipidis, E. and Gesbert, D. and Haardt, M. and Ho, Ka-Ming. an Jorswieck, E.
%and Larsson, Erik G. and Li, Jianhui and Lindblom, J. and Scheunert, C. and Schubert, M. and Vucic, N.},
%journal={in Proc. Future Network and Mobile Summit Conference},
%title={Transmit Beamforming for Inter-Operator Spectrum Sharing},
%year={2011},
%}
%
%
%
%@ARTICLE{Larsson2008,
%author={Larsson, E.G. and Jorswieck, E.A.},
%journal={IEEE J. Sel. Areas Commun.},
%title={Competition versus Cooperation on the MISO Interference Channel},
%volume={26},
%month = {Aug.},
%year={2008},
%number={7},
%pages={1059-1069},
%}
%
%
%@ARTICLE{Perea2013,
%author={Perea-Vega, D. and Girard, A. and Frigon, J.F.},
%journal={EURASIP Journal on Wireless Communications and Networking},
%title={Dual-based bounds for resource allocation in zero-forcing beamforming OFDMA-SDMA systems},
%volume={},
%month = {Feb.},
%year={2013},
%number={},
%pages={},
%}
%
%
%@ARTICLE{Wiesel2008,
%author={Wiesel, A. and Eldar, Y.C. and Shamai, S.},
%journal={IEEE Trans on Signal Processing},
%title={Zero-Forcing Precoding and Generalized Inverses},
%year={2008},
%month={Sept},
%volume={56},
%number={9},
%pages={4409-4418},
%}
%
%
%
%
%
%@ARTICLE{WeiYu2007,
%author={Wei Yu and Tian Lan},
%journal={IEEE Trans on Signal Processing},
%title={Transmitter Optimization for the Multi-Antenna Downlink With Per-Antenna Power Constraints},
%year={2007},
%volume={55},
%number={6},
%pages={2646-2660},
%}
%
%
%
%
%
%@ARTICLE{Scaglione2002,
%author={Scaglione, A. and Stoica, Petre and Barbarossa, S. and Giannakis, G.B. and Sampath, H.},
%journal={IEEE Transactions on Signal Processing},
%title={Optimal designs for space-time linear precoders and decoders},
%year={2002},
%volume={50},
%number={5},
%pages={1051-1064},
%}
%
%@ARTICLE{Coiffi,
%author={Christensen, S.S. and Agarwal, R. and Carvalho, E. and Cioffi, J.M.},
%title={Weighted sum-rate maximization using weighted MMSE for MIMO-BC beamforming design},
%journal={IEEE Trans. on Wireless Commun.},
%year={2008},
%volume={7},
%number={12},
%pages={4792-4799},
%}


\end{thebibliography}

\end{document}